\documentclass{article}
\usepackage{graphicx} 
\usepackage{geometry}
\usepackage{array}
\usepackage{multirow}
\usepackage{authblk}
\usepackage{setspace}
\usepackage{lineno}
\usepackage[colorlinks=true, urlcolor=blue, linkcolor=red]{hyperref}

\usepackage[
backend=biber,
style=authoryear,
citestyle=apa,
sorting=nyt
]{biblatex}

\title{Bibliography management: \texttt{biblatex} package}

\title{Towards Risk Analysis of the Impact of AI on the Deliberate Biological Threat Landscape}
\author{Matthew E. Walsh \\
\texttt{mwalsh52@jhu.edu}}
\medskip
\affil{Department of Environmental Health and Engineering}
\affil{Johns Hopkins Bloomberg School of Public Health}
\date{June 10, 2024}
\begin{document}

\maketitle

\begin{abstract}
    The perception that the convergence of biological engineering and artificial intelligence (AI) could enable increased biorisk has recently drawn attention to the governance of biotechnology and artificial intelligence. The 2023 Executive Order, \textit{Executive Order on the Safe, Secure, and Trustworthy Development and Use of Artificial Intelligence}, requires an assessment of how artificial intelligence can increase biorisk. Within this perspective, quantitative and qualitative frameworks for evaluating biorisk are presented. Both frameworks are exercised using notional scenarios and their benefits and limitations are then discussed. Finally, the perspective concludes by noting that assessment and evaluation methodologies must keep pace with advances of AI in the life sciences. 
\end{abstract}
\section{Introduction}

Biological weapons are a combination of a biological agent and a mechanism to disseminate the agent to a target population. Many \href{https://www.dfat.gov.au/publications/minisite/theaustraliagroupnet/site/en/human_animal_pathogens.html}{pathogens and toxins} can be weaponized and include pathogens that cause Anthrax, Plague, Smallpox, Tularemia, and Viral hemorrhagic fevers (e.g., Ebola Virus Disease), and the toxin that causes Botulism. The causative agents of these diseases can easily infect large numbers of people because they are person-to-person transmissible or easily disseminated over large areas. These diseases have high mortality rates, would cause public panic, and disrupt day-to-day life.

There is currently an international norm that the use of biological weapons is unacceptable. This norm is codified through the Biological Weapons Convention (BWC), an international treaty. The BWC, which entered into force in 1975, bans Nation States from the development, production, acquisition, transfer, and stockpiling of biological weapons. The BWC is currently signed by 185 Nation States and is complemented by UN Security Resolution 1540 and the Geneva Convention. Resolution 1540 requires all States to prevent the proliferation of biological weapons to non-state groups and the Geneva Convention prohibits use of biological weapons. However, the 1984 \textit{Salmonella} salad bar attacks by the Rajneesh in the United States, the 1993 anthrax attack attempts by the Aum Shinrikyo cult in Japan, as well as the Amerithrax letter attacks of 2001, serve as reminders that policy measures can only go so far in preventing the misuse of biology. Today, the United States assesses that China and Iran display activities that raise concern regarding biological weapons and that North Korea and Russia maintain active biological weapons programs (\cite{us_department_of_state_adherence_2023}).

Around the time the BWC was entering into force, researchers were meeting in Pacific Grove, California to discuss the implications of an important new advance – recombinant DNA technology (\cite{berg_asilomar_2008}). Researchers had recently demonstrated the ability to put DNA from one organism into another, and the scientific community had concerns about the human and environmental health risks in addition to optimism about its potential benefits. Scientists called for a moratorium on such work until the magnitude of risks were assessed and stringent safety guidelines were established. The meeting, now simply known as the Asilomar Conference, named after the conference center, paved the way for responsible governance of recombinant DNA technology. Today, many dual use technologies (those that can be used for both beneficial and malicious purposes) present governance challenges because actions that would mitigate their risks would likely restrict their potential benefits.

The ways in which dual use life science technologies could be misused have been categorized by the late Jonathan B. Tucker and include, 1) “facilitate or accelerate the production of standard biological warfare agents”, 2) “identify or develop novel biological warfare agents”, 3) “result in new knowledge that could be misused to develop a new or enhanced means of harming the human body”, and 4) “lead to harmful applications that undermine international legal norms” (\cite{jonathan_b_tucker_innovation_2012}). Since the Asilomar conference in 1975, many technological developments beyond recombinant DNA have demonstrated the potential to be misused and are therefore deemed to have “dual-use potential”. Such technologies include protein engineering, synthesis of viral genomes, gene drives, personal genomics, gene therapy, CRISPR, and immunological modulation. In addition to their benefits, these technologies make it easier for researchers to engineer organisms in ways that make them more suitable to cause harm. A 2004 NASEM report, known as The Fink Report, details seven experimental outcomes that are particularly concerning (\cite{committee_on_research_standards_and_practices_to_prevent_the_destructive_application_of_biotechnology_national_research_council_biotechnology_2004}). Building upon The Fink Report, the 2018 NASEM report Biodefense in the Age of Synthetic Biology includes a framework for evaluating the relative concern of synthetic biology enabled capabilities (\cite{committee_on_strategies_for_identifying_and_addressing_potential_biodefense_vulnerabilities_posed_by_synthetic_biology_biodefense_2018}). More recently, the perception that the convergence of biological engineering and artificial intelligence (AI) could enable increased harms has drawn attention to the governance of biotechnology and artificial intelligence (\cite{helena_foundation_biosecurity_2023,sarah_r_carter_convergence_2023}). 

\section{Biosecurity Risk}

President Biden recently signed the \textit{Executive Order on the Safe, Secure, and Trustworthy Development and Use of Artificial Intelligence}. Section 4.4.ii.a of the Executive Order requires the Secretary of Defense, in coordination with many others, to complete a study that, “assesses the ways in which AI can increase biosecurity risks, including risks from generative AI models trained on biological data, and makes recommendations on how to mitigate these risks” (\cite{white_house_executive_2023}). Within this perspective, a quantitative framework and qualitative process are proposed and discussed for assessing the ways in which AI can increase biosecurity risks. 

Biosecurity risk can be defined as a complex scenario in which a biological organism or toxin is intentionally used to cause harm or damage. For the purposes of this work, it is helpful to consider a risk quantification framework as the triplet ($S_{i}$, $P_{i}$, $C_{i}$), where $S_{i}$ is the \textit{i}th scenario, $P_{i}$ is the probability of that scenario occurring, and $C_{i}$ is the consequence of the \textit{i}th scenario, where \textit{i} = 1,2,…N (\cite{aven_society_2018}). To enable comparison, the use of artificial intelligence is assumed to create a parallel set of \textit{ai} scenarios with probabilities and consequences, denoted ($S_{ai}$, $P_{ai}$, $C_{ai}$), for each of the \textit{N} scenarios. Each \textit{ai} scenario can be considered an \textit{i} scenario that has been modified to include the use of AI. Assuming the biosecurity risk of each of the \textit{i} scenarios is discrete, AI will increase biosecurity risk when there is an increase in the probability or consequence term of a given \textit{ai} scenario relative to the corresponding \textit{i} scenario. Succinctly, AI can increase biosecurity risk through increasing the probability and/or consequence of using biology to cause harm or damage, assuming the paired term (i.e., probability or consequence) does not decrease more than the increase.

\section{Probability}

For the purposes of this perspective, each scenario is considered to be composed of five steps, denoted $s_{i}^{step}$ or $s_{ai}^{step}$.  Each of the five steps is described in Table 1 and has an associated probability, $p_{i}^{step}$ or $p_{ai}^{step}$. Use of a biorisk chain model to describe all potential misuse scenarios may undervalue or oversimplify components of a misuse scenario. In addition, scenarios of misuse may or may not include all steps in a given biorisk chain model. However, establishing and using biorisk chain models in biorisk analysis enables evaluation of the impact of AI on biorisk and clear communication around misuse steps, underlying assumptions, and components of the analysis (\cite{sandberg_who_2020}). At present, $S_{i}$ and $S_{ai}$ will have the same steps because AI tools are currently not at a maturity in which they are able to reliably replace a step or obviate the need of a future step. 

\begin{table}[b]
    \centering
\caption{Component steps of a biorisk chain}
\label{tab:my_label}
    \begin{tabular}{|m{9em}|m{25em}|m{5em}|} \hline 
        \textbf{Step} & \textbf{Description}  & \textbf{Notation}\\ \hline 
        Ideation & The conceptualization of a plan to cause harm and involving the use of a biological or toxin weapon & $s_{i}^{idea}$ \\ \hline 
        Acquisition & Actions related to acquiring the pathogen, toxin, delivery mechanism and/or any other necessary components  & $s_{i}^{acq}$\\ \hline 
        Production & Actions related to producing the pathogen or toxin to a desired quantity & $s_{i}^{prod}$\\ \hline 
        Weaponization & Actions related to modifying a pathogen or toxin, post-production, to make it compatible with its intended use as a weapon, including combining the biological agent with a delivery device  & $s_{i}^{weapon}$\\ \hline 
        Deploy/Delivery & Actions related to using a biological weapon against a target &$s_{i}^{deploy}$ \\ \hline 
    \end{tabular}   
\end{table}

The probability term, $P_{i}$ and $P_{ai}$, is considered to represent the likelihood of successful execution of all steps of a sequential plan, spanning Ideation through Deploy/Deliver, and regardless of the consequence. For each step, the probability of success can be influenced by many factors including the required resources, skills, and knowledge to perform the step. Probability of success can also be influenced by external policy factors like export controls or DNA synthesis screening regulations. Consequence can be influenced by the steps taken in the plan itself, but also by external factors like preparedness to respond to a plan (e.g., with vaccines or antibiotics). Consequence can also be influenced by human behavior (e.g., mask wearing), policy decisions (e.g., mask mandates), and natural phenomena (e.g., weather). As such, consequence cannot be determined by the characteristics of the pathogen and delivery method alone. Finally, success and consequence can be influenced by the decision to include or exclude steps from the plan.  

For example, if a nefarious actor intended to spread \textit{Salmonella} at a salad bar to prevent community members from voting in a local election, the probability and consequence terms may be altered based on the specific strain of \textit{Salmonella} used. A more pathogenic strain of \textit{Salmonella} may be more challenging to acquire, therefore decreasing the probability of successfully executing the overall plan because of a barrier introduced at one step of the plan. At the same time, this decision would increase the consequence term corresponding to severity of disease. But because the disease would be more severe or unusual, it may be more likely to generate a more significant public health response that could decrease the consequence. If the strain of \textit{Salmonella} were well-characterized, the uncertainties of the consequences would be well understood. However, if the strain were less well-characterized in the environment, perhaps because it was engineered, the uncertainty of the consequence would increase. Given the many component parameters of the probability and consequence terms, they are not independent of each other and will inherently have meaningful associated uncertainty. Therefore, simplifying assumptions to allow only one component of the scenario pair to change may be required to perform a comparison.

This framework can be exercised to identify where AI can increase biosecurity risk. Taking a simplified scenario $S_{i}$ in which a relatively knowledgeable lone actor has an insight to synthesize a pathogen in a lab using nucleic acid material provided by a mail order DNA synthesis company. The specific bottleneck to consider in this comparison is that many mail-order DNA synthesis companies evaluate incoming orders to prevent malicious actors from acquiring potentially harmful material. To overcome this bottleneck, an actor would need to identify a company that would provide them the material – and for this scenario comparison it is assumed that the actor has a limited and fixed amount of time. Within $s_{i}^{idea}$, it can be assumed that the actor uses Google or another search engine. In the comparator scenario step, $s_{ai}^{idea}$, the lone actor uses an LLM-based chatbot and Google. If the outcome of this step is assumed to be simply “success” or “fail”, there are then four paired outcomes to consider. To enable a simplistic comparison, “success” is assumed to be reached in the same way (i.e., identifying the same company to order from), to allow for the difference between future $s_{i}$ and $s_{ai}$ to remain negligible. 

Given the general understanding that AI-powered chatbots make it easier to acquire information that is available online, and that the information needed to overcome the screening bottleneck is available online, it is reasonable to assume $p_{ai}^{idea}$ is greater than $p_{i}^{idea}$ and that AI-powered chatbots therefore increase biosecurity risk in this way. But because the magnitude of the change is not known, it is challenging for a risk manager to evaluate if the increase in risk is meaningful or acceptable. If each $p_{i}^{idea}$ and $p_{ai}^{idea}$ were quantified, a difference in risks contributed by the scenario pairs of (success | fail) and (fail | success) could be estimated. 

Two similar research efforts have attempted to quantify the contribution of LLM-based chatbots on the ability to successfully develop a plan to misuse biology, considered $p^{idea}$ within the framework discussed here. The first effort was conducted by the RAND Corporation and the second was by OpenAI (\cite{mouton_operational_2023, tejal_patwardhan_building_2024}). Both research teams found no statistical difference between $p_{i}^{idea}$ and $p_{ai}^{idea}$, although each team identified multiple methodological improvements that could be made to strengthen the confidence in their results.

Within the OpenAI exercise, participants were asked to complete five tasks that "were designed to assess the end-to-end critical knowledge needed to complete each stage in the biological threat creation process."(\cite{tejal_patwardhan_building_2024}). Each of the five tasks roughly corresponds to one step in the biorisk chain defined in Table \ref{tab:my_label}. Half of the 100 participants (n = 50) were given access to an LLM-based chatbot and the internet, while the other half (n = 50) only had access to the internet. Participants generated text-based outputs that were then evaluated by biosecurity experts for accuracy and completeness, among other performance metrics, and the performance metrics were compared between those with access to an LLM-based chatbot and those without access to an LLM-based chatbot.

\begin{figure}
    \centering
    \includegraphics[width=1\linewidth]{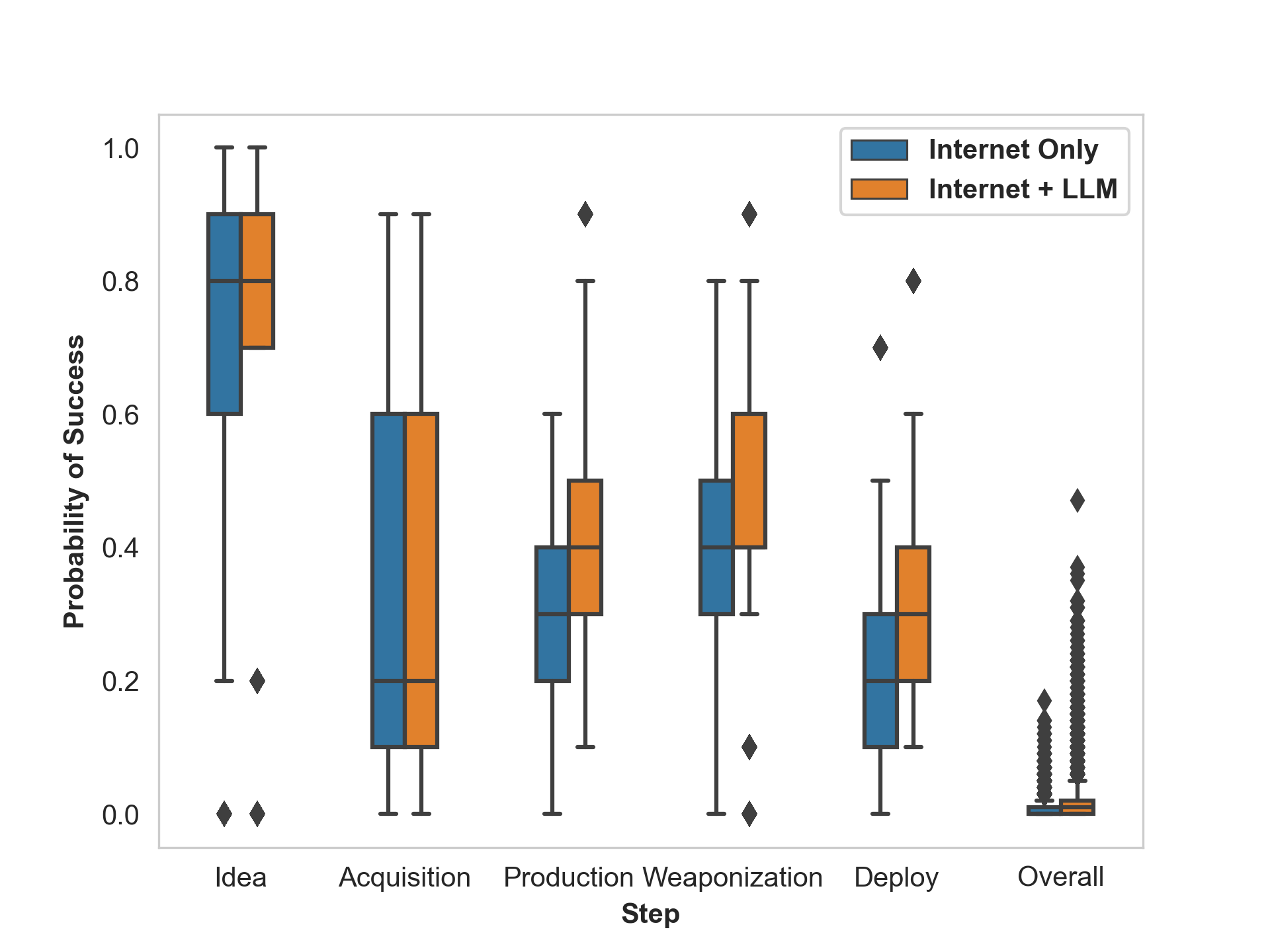}
    \caption{Box-and-whisker plot displaying results of 10,000 Monte Carlo simulations of the probability of successfully developing a plan for the misuse of biology. Data distributions are based on data from OpenAI (\cite{tejal_patwardhan_building_2024}). Blue results represent participants with access to the internet only. Orange results represent participants with access to the internet and an LLM-based chatbot. Boxes represent the median and inter-quartile range (IQR). Diamonds denote outliers more than 1.5x the IQR from Q1 or Q3. "Overall" results represent the product of the probability of each individual step.}
    \label{fig:1}
\end{figure}

The biosecurity experts scored the accuracy of responses on a scale of 0 to 10; a 10 represented that the participant included all key steps needed to complete the task. The triplet risk quantification framework proposed above can be exercised using the data OpenAI has shared. Recognizing that the ability to develop a plan and to execute it are distinct, the accuracy scores from the OpenAI data can serve as a proxy for $p$, and the contribution to risk can be computed. Herein, a Monte Carlo simulation was conducted using the distribution of accuracy scores from the OpenAI work as the probability distributions from which each $p^{step}$ was sampled (\ref{fig:1}, \ref{tab:2}). The simulation was run 10,000 times, with the overall probability calculated as the product of the probability of each step, and executed in Python. Given that participants were asked to complete the same five tasks, it is assumed that the consequence of $s_{i}$ and $s_{ai}$ are equal, notionally and for simplicity defined as 100,000 deaths. By multiplying the probability and consequence terms, the biosecurity risk is determined to be 500 deaths for $s_{i}$ and 1,700 deaths for $s_{ai}$. Therefore, in this theoretical pair of scenarios, access to an LLM-based chatbot would be considered to increase biosecurity risk by 1,200 deaths.

Risk managers should consider more than the magnitude of the increase in biorisk when determining if such an increase is acceptable. Increases in risk attributable to a specific technology should be considered in the context of the benefits of that technology as well. This relative comparison relies on quantifying the benefits of the technology, which is a similarly complex task to quantifying risk. Further, consideration should be given to societal and ethical norms. 

\begin{table}[b]
    \centering
\caption{Monte Carlo Simulation Results}
\label{tab:2}
    \begin{tabular}{|m{5em}|m{10em}|m{10em}|m{10em}|} \hline 
        \textbf{Scenario} & \textbf{Average Overall $P$}  & \textbf{Notional $C$ (deaths)} & \textbf{Risk (deaths per scenario)}\\ \hline 
        $s_{i}$ & 0.005 & 100,000 & 500 \\ \hline 
        $s_{ai}$ & 0.017 & 100,000 & 1700 \\ \hline 
    \end{tabular}   
\end{table}

\section{Consequence}
The above scenario comparison is somewhat tractable because the consequence term remains fixed. An increase in the consequence term would also result in an increase in biosecurity risk, assuming that the probability term does not decrease commensurately or more. However, such an assumption is likely unwarranted given the current state of AI technologies in the life sciences and biological engineering. A commonly stated concern is that AI tools may be able to design pathogens that are more transmissible and more pathogenic than those found in nature (\cite{sandbrink_artificial_2023}). These characteristics of a pathogen are often used as a proxy for consequence. However, as alluded to in the salad bar scenario above, the public health response may differ depending on the severity of disease, altering the consequence. Therefore, assessments about the consequence of a deliberate biological attack should be accompanied by a significant amount of uncertainty because it is dependent on unknowable human behavior. Furthermore, this scenario assumes that AI-assisted engineering methods will be able to reach a set of viable end-states (e.g., pathogen design) that are not included within the set of viable end-states of traditional biological engineering methods. To date, this assumption remains largely unproven and lacks adequate rationale. Given the ability to explicitly synthesize DNA sequences, there is no reason to think that an AI tool would generate a design that could not theoretically be designed or identified with sufficient rounds of traditional engineering, such as random mutagenesis. For comparison purposes, and until proven otherwise, it is therefore likely better to assume that traditional engineering approaches and AI-assisted engineering approaches are able to reach outcomes of similar consequence. 

If, for the sake of comparison, it is assumed that both $S_{i}$ and $S_{ai}$ have the same set of consequences, the impact of purpose-built AI-based engineering tools on the $p_{ai}$ of each $s_{ai}$ compared to the $p_{i}$ of each $s_{i}$ should be explored, similar to the evaluation of the impact of the chatbots in the prior example. Consider a scenario in which a nefarious actor wants to engineer a pathogen using existing bioengineering methods and the paired scenario containing the use of AI. In this scenario pair, it is assumed that the use of genetic engineering within the acquisition step, that requires a traditional synthetic biology process of design – build – test – learn and with the end goal of optimizing the characteristic properties of a pathogen (not the pathogen’s effects on its host), is used (\cite{khalil_synthetic_2010}). Within this scenario, it is assumed that there is a biomolecule, or “candidate sequence”, that has desirable properties that could be optimized. In $s_{i}^{acq}$ it is assumed the use of random mutagenesis or equivalent method to identify the variants of the candidate sequence to build and test, whereas in $s_{ai}^{acq}$ an AI system generates the recommended sequences to build and test. For the purposes of this theoretical evaluation, it is assumed that the requirements (e.g., resources, knowledge) to perform mutagenesis and AI design are equal (although, in practice this is likely not true). In this $s_{i}$ and $s_{ai}$ pair, the build, test, and learn processes are equivalent because they are independent of the AI tool under evaluation. Notably, a scenario in which AI is used in both the design and learn processes is quite feasible, however this would result in a distinct $s_{ai}^{acq}$. Therefore, the contribution of AI comes from potentially decreasing the overall number of design-build-test-learn cycles which would reduce the time, cost, and resources required to achieve the desired end goal. To support risk analysis, an understanding of the difference in number of design-build-test-learn cycles between a typical $S_{i}$ and $S_{ai}$ (as a means to characterize a difference in probability of success, with decreasing probability associated with increasing numbers of cycles) is required.

Currently, there is not enough information in the published literature to support a reasonable, generalized characterization of this difference. Often, when researchers are developing AI systems and tools for engineering biology, they will perform the same number of design-build-test-learn cycles to demonstrate that the AI method is improved over the traditional engineering method. For example, a previously reported machine learning method provided a nearly 29-fold improvement compared to a traditional method (\cite{li_machine_2023}). However, it took the team 12+ months to develop the machine learning method that provided the designs. If accounting for the time (and cost, knowledge, resources, etc.) a more apt comparison may be to the number of traditional engineering cycles that could have occurred over that 12+ month development period. Additionally, the authors tested their methods in two very similar engineering tasks and notably, the method that resulted in the best performance in the first scenario gave the worst performance in the second scenario without improving over the traditional engineering method. Lastly, the authors did not perform additional traditional engineering cycles to determine the number of cycles it would take to reach performance of the best AI outcome. As AI tools are increasingly used in academic and industry settings, time and cost savings could be quantified to provide value in such characterizations. However, to fully understand the impact of including AI methods and tools in biological engineering processes in a manner that supports risk assessment, dedicated efforts to characterize the performance of AI methods across diverse tasks is needed.

Efforts to perform characterization of the ability of AI tools to assist in the engineering of pathogens are likely to be controversial. Given that the output of an AI-enabled bioengineering tool is a prediction, validation of the prediction in a biological laboratory is necessary to determine how often the predictions are correct and if there are circumstances or conditions, such as amount or type of training data, that influence the rate of success. Unlike chemical threats, biological threats can transmit from person-to-person and self-amplify. As such, a small mistake in a laboratory could result in large-scale consequence. With this in mind, there should be limits to the extent of characterization of the ability of AI tools to assist in the engineering of pathogens, including in the creation of novel pathogens, for the sole purpose of informing risk analysis. Creating and evaluating biological agents that are predicted to be highly pathogenic should not be done for the sole purpose of characterizing the predictive capabilities of these tools. This should be the norm.

Proxy evaluation metrics for the ability of AI tools to deliberately design novel pathogens could offer a possible path forward. However, it may be challenging for the life science, biosecurity and associated policy communities to agree on sufficient proxy metrics given longstanding disagreements about the ability to “reasonably anticipate” the outcome of scientific experiments (\cite{goodrum_virology_2023}). Recent updates to United States Government policy include a definition of "reasonably anticipated", but it remains to be seen if this definition will help resolve these issues (\cite{office_of_science_and_technology_policy_united_2024}). While increased use of AI tools and collection of data around their use over time may help inform expected outcomes, it may be better to consider the capability of an AI tool as a clearly stated assumption in risk analysis conversations or to seek a relative understanding of the level of concern associated with capabilities and scenarios in which biology is used to cause harm. The notable number of assumptions required to use the triplet framework described herein may restrict its broad utility in quantitative assessments, but nonetheless should provide a valuable basis for discussion around the impact of AI on the deliberate biological threat landscape.

\section{Qualitative Risk}
An alternative approach to determining the acceptability of an increase in risk associated with the impact of AI on a given step in the scenario could be to qualitatively evaluate the relative contribution of each step to the overall scenario and then compare this evaluation with the potential benefits of the AI technology. In the biorisk chain, each step has different limiting factors making the relative contribution of $p_{i}$ to $P_{i}$ variable (Table 2). Given that the magnitude of the risk is not known, experts are often relied upon to conceptualize a meaning of “low/med/high”. This could then allow for an assessment of if the impact of AI on $s_{i}$ results in an $s_{ai}$ with an increased or decreased $p_{ai}$ compared to $p_{i}$, potentially changing the categorical contribution to $S_{i}$.

\begin{table}[b]
    \centering
\caption{Notional requirement levels for completion of each step within the biorisk chain by an unskilled actor and associated probability. Note: The entries in this table are judgements made by the author alone.  It is not intended to serve as an authoritative source on requirement levels for each step.}

\label{tab:3}
    \begin{tabular}{|c|c|c|c|c|c|c|} \hline 
        \textbf{Step} & \textbf{Time} & \textbf{Cost} & \textbf{Knowledge} & \textbf{Resources} & \textbf{Safeguard} & \textbf{Relative $p_{i}^{step}$}\\ \hline 
        Ideation & Low & Low & Low & Low & Low & High \\ \hline 
        Acquisition & Low & Low & Low & Low & Med & Low \\ \hline 
        Production & High & High & High & High & High & Low\\ \hline 
        Weaponization & High & High & High & High & High & Low\\ \hline 
        Deploy/Delivery & Low & High & High & Med & High & Med\\ \hline
    \end{tabular}

\end{table}

In this biorisk paradigm, concerning impacts of AI would likely result in changes of $p_{ai}$ from “low” to “medium” or “high” and from “medium” to “high”. Given the notional content of Table 2, impacts of AI on the acquisition, production, and weaponization steps may be more important to understand and analyze than those at the ideation step.

The NASEM Report \textit{Biodefense in the Age of Synthetic Biology} offers a viable starting point for performing a qualitative risk assessment as required by the recent Executive Order. The report provides relative levels of concern for a set of synthetic biology enabled capabilities; these capabilities will be impacted by artificial intelligence tools, some more so than others. Of particular relevance today are the capabilities of making existing bacteria and viruses more dangerous. The NASEM report includes a list of nine viral traits and eight bacterial traits that could be engineered to result in a more dangerous pathogen. With this in mind, the following steps could be taken to update the evaluations within the NASEM report:
\begin{enumerate}
    
\item Identify organisms of interest. Organisms may biologically differ in the way in which their traits of interest are determined. As such, the way these traits are engineered may be different and thus the AI-enabled tools used to support engineering may be different. This list of organisms could include pathogens with pandemic potential, those that are likely candidates for misuse (e.g., Australia Group List), and those that are often engineered in beneficial biotechnological applications because they likely have existing AI-enabled engineering tools.

\item Identify existing AI-enabled tools that are amenable to engineering the traits of the organisms of interest through literature, web search, and key informant interviews. Some tools may be purpose-built for engineering whereas other tools may be designed to characterize organisms and in some scenarios could be repurposed or modified to engineer the characterized trait.

\item When AI-enabled tools do exist, systematically and qualitatively evaluate their impact on the four categories that, collectively, contribute to the level of concern that should be attributed to a particular scenario. The four categories are Usability of the Technology, Usability as a Weapon, Requirements of Actors, and Potential for Mitigation, with each having multiple sub components. Risk assessors should seek to understand the common uses of the AI tools and how the tools impact the sub components, such as ease of use or required expertise.  

\item Given the findings of Step 3, review and update, if necessary, the overall relative level of concern for each capability of concern in the NASEM report.

\end{enumerate}

Similar processes could also be developed and executed for assessing the impact of AI on the capabilities that do not include the engineering of pathogens, such as those that modify the human microbiome, immune system or genome. In aggregate, the outcome of this potential work would be a relative comparison of levels of concern about the capabilities. It would not necessarily allow for a conclusion about overall changes in the absolute level of biorisk, but rather an understanding of how AI impacts potentially concerning scenarios in a relative manner. Even without an absolute quantitative understanding of biorisk, a relative understanding helps risk managers devote the appropriate attention and resources to mitigate biorisk.

\section{Moving Forward}

As AI-enabled biological engineering tools continue to mature, it will be increasingly important to understand the ways in which they impact the deliberate biological threat landscape. Such advances include retrieval-augmented generation (a technique that supplements an LLM's internal knowledge with external, predefined sources), the development and release of multi-modal models, and the ability of LLMs to access the internet and other software tools, including other AI tools purpose built for life science applications. (\cite{lala_paperqa_2023, bran_chemcrow_2023}). These ongoing efforts have started to blur the distinction between LLM-based chatbots and purpose-built AI tools for engineering biology. Methodologies for evaluating the impacts of AI on all steps of the biorisk chain will need to keep pace with these advances and likely require dedicated laboratory resources. 

 Given the complexity of the misuse of biology, risk managers and risk assessors will need to work together to clearly articulate their goals and assumptions within their work. Quantitative and qualitative risk analysis, informed by a technical understanding of the ways in which AI tools are used within biology and the life sciences, will be useful, each with different limitations and benefits. The technical understanding of how AI is used and performs within varied life science scenarios can be furthered by dedicated characterization efforts of AI-enabled tools outside of pathogen engineering context. This work, however, is noticeably absent at present. 

\section{Acknowledgements}

The author thanks Dr. Gerald Epstein and Dr. Gigi Kwik Gronvall for their constructive comments and suggestions on the manuscript. The author is funded by Open Philanthropy. 

\medskip

\printbibliography

@misc{office_of_science_and_technology_policy_united_2024,
	title = {United {States} {Government} {Policy} for {Oversight} of {Dual} {Use} {Research} of {Concern}  and {Pathogens} with {Enhanced} {Pandemic} {Potential}},
	url = {https://www.whitehouse.gov/wp-content/uploads/2024/05/USG-Policy-for-Oversight-of-DURC-and-PEPP.pdf},
	author = {{Office of Science and Technology Policy}},
	month = may,
	year = {2024},
}

@misc{tejal_patwardhan_building_2024,
	title = {Building an early warning system for {LLM}-aided biological threat creation},
	url = {https://openai.com/index/building-an-early-warning-system-for-llm-aided-biological-threat-creation/},
	urldate = {2024-05-09},
	author = {{Tejal Patwardhan} and {Kevin Liu} and {Todor Markov} and {Neil Chowdhury} and {Dillon Leet} and {Natalie Cone} and {Caitlin Maltbie} and {Joost Huizinga} and {Carroll Wainwright} and {Shawn (Froggi) Jackson} and {Steven Adler} and {Rocco Casagrande} and {Aleksander Madry}},
	month = jan,
	year = {2024},
}

@techreport{mouton_operational_2023,
	title = {The {Operational} {Risks} of {AI} in {Large}-{Scale} {Biological} {Attacks}: {A} {Red}-{Team} {Approach}},
	shorttitle = {The {Operational} {Risks} of {AI} in {Large}-{Scale} {Biological} {Attacks}},
	url = {https://www.rand.org/pubs/research_reports/RRA2977-1.html},
	language = {en},
	urldate = {2024-02-07},
	institution = {RAND Corporation},
	author = {Mouton, Christopher A. and Lucas, Caleb and Guest, Ella},
	month = oct,
	year = {2023},
}

@book{committee_on_research_standards_and_practices_to_prevent_the_destructive_application_of_biotechnology_national_research_council_biotechnology_2004,
	address = {Washington, D.C.},
	title = {Biotechnology {Research} in an {Age} of {Terrorism}},
	isbn = {978-0-309-08977-7},
	url = {http://www.nap.edu/catalog/10827},
	urldate = {2023-10-25},
	publisher = {National Academies Press},
	author = {{Committee on Research Standards and Practices to Prevent the Destructive Application of Biotechnology, National Research Council}},
	month = feb,
	year = {2004},
	doi = {10.17226/10827},
}

@misc{white_house_executive_2023,
	title = {Executive {Order} on the {Safe}, {Secure}, and {Trustworthy} {Development} and {Use} of {Artificial} {Intelligence}},
	url = {https://www.whitehouse.gov/briefing-room/presidential-actions/2023/10/30/executive-order-on-the-safe-secure-and-trustworthy-development-and-use-of-artificial-intelligence/},
	language = {en-US},
	urldate = {2023-12-18},
	journal = {The White House},
	author = {White House, The},
	month = oct,
	year = {2023},
}

@techreport{helena_foundation_biosecurity_2023,
	title = {Biosecurity in the {Age} of {AI}: {Chairperson}'s {Statement}},
	url = {https://www.helenabiosecurity.org/},
	institution = {Helena Foundation},
	author = {{Helena Foundation}},
	month = jul,
	year = {2023},
}

@techreport{sarah_r_carter_convergence_2023,
	title = {The {Convergence} of {Artificial} {Intelligence} and the {Life} {Sciences}},
	url = {https://www.nti.org/analysis/articles/the-convergence-of-artificial-intelligence-and-the-life-sciences/},
	language = {en},
	urldate = {2024-01-22},
	author = {{Sarah R. Carter} and {Nicole Wheeler} and {Sabrina Chwalek} and {Chris Isaac} and {Jaime M. Yassif}},
	month = oct,
	year = {2023},
}

@techreport{aven_society_2018,
	title = {Society for risk analysis glossary},
	url = {https://www.sra.org/wp-content/uploads/2020/04/SRA-Glossary-FINAL.pdf},
	urldate = {2024-01-22},
	institution = {Society for Risk Analysis},
	author = {Aven, Terje and Ben-Haim, Yakov and Boje Andersen, Henning and Cox, Tony and Droguett, Enrique López and Greenberg, Michael and Guikema, Seth and Kröger, Wolfgang and Renn, Ortwin and Thompson, Kimberly M},
	year = {2018},
}

@article{khalil_synthetic_2010,
	title = {Synthetic biology: applications come of age},
	volume = {11},
	issn = {1471-0056},
	number = {5},
	journal = {Nature Reviews Genetics},
	author = {Khalil, Ahmad S and Collins, James J},
	year = {2010},
	note = {Publisher: Nature Publishing Group UK London},
	pages = {367--379},
}

@misc{lala_paperqa_2023,
	title = {{PaperQA}: {Retrieval}-{Augmented} {Generative} {Agent} for {Scientific} {Research}},
	shorttitle = {{PaperQA}},
	url = {http://arxiv.org/abs/2312.07559},
	doi = {10.48550/arXiv.2312.07559},
	urldate = {2023-12-14},
	publisher = {arXiv},
	author = {Lála, Jakub and O'Donoghue, Odhran and Shtedritski, Aleksandar and Cox, Sam and Rodriques, Samuel G. and White, Andrew D.},
	month = dec,
	year = {2023},
	note = {arXiv:2312.07559 [cs]},
}

@article{berg_asilomar_2008,
	title = {Asilomar 1975: {DNA} modification secured},
	volume = {455},
	copyright = {2008 Springer Nature Limited},
	issn = {1476-4687},
	shorttitle = {Asilomar 1975},
	url = {https://www.nature.com/articles/455290a},
	doi = {10.1038/455290a},
	language = {en},
	number = {7211},
	urldate = {2023-10-25},
	journal = {Nature},
	author = {Berg, Paul},
	month = sep,
	year = {2008},
	note = {Number: 7211
Publisher: Nature Publishing Group},
	pages = {290--291},
}

@techreport{us_department_of_state_adherence_2023,
	title = {Adherence to and {Compliance} {With} {Arms} {Contorl}, {Nonproliferation}, {And} {Disarmament} {Agreements} and {Commitments}},
	url = {https://www.state.gov/wp-content/uploads/2023/04/13APR23-FINAL-2023-Treaty-Compliance-Report-UNCLASSIFIED-UNSOURCED.pdf},
	urldate = {2023-10-25},
	author = {{U.S. Department of State}},
	month = apr,
	year = {2023},
	pages = {22--28},
}

@book{jonathan_b_tucker_innovation_2012,
	address = {Cambridge, Massachusetts},
	title = {Innovation, {Dual} {Use}, and {Security}},
	publisher = {The MIT Press},
	author = {{Jonathan B. Tucker}},
	year = {2012},
}

@misc{sandbrink_artificial_2023,
	title = {Artificial intelligence and biological misuse: {Differentiating} risks of language models and biological design tools},
	shorttitle = {Artificial intelligence and biological misuse},
	url = {http://arxiv.org/abs/2306.13952},
	doi = {10.48550/arXiv.2306.13952},
	urldate = {2023-10-03},
	publisher = {arXiv},
	author = {Sandbrink, Jonas B.},
	month = aug,
	year = {2023},
	note = {arXiv:2306.13952 [cs]},
}

@misc{bran_chemcrow_2023,
	title = {{ChemCrow}: {Augmenting} large-language models with chemistry tools},
	shorttitle = {{ChemCrow}},
	url = {http://arxiv.org/abs/2304.05376},
	doi = {10.48550/arXiv.2304.05376},
	urldate = {2023-09-06},
	publisher = {arXiv},
	author = {Bran, Andres M. and Cox, Sam and White, Andrew D. and Schwaller, Philippe},
	month = jun,
	year = {2023},
	note = {arXiv:2304.05376 [physics, stat]},
}

@article{goodrum_virology_2023,
	title = {Virology under the {Microscope}—a {Call} for {Rational} {Discourse}},
	volume = {97},
	url = {https://journals.asm.org/doi/full/10.1128/jvi.00089-23},
	doi = {10.1128/jvi.00089-23},
	number = {2},
	urldate = {2023-08-23},
	journal = {Journal of Virology},
	author = {Goodrum, Felicia and Lowen, Anice C. and Lakdawala, Seema and Alwine, James and Casadevall, Arturo and Imperiale, Michael J. and Atwood, Walter and Avgousti, Daphne and Baines, Joel and Banfield, Bruce and Banks, Lawrence and Bhaduri-McIntosh, Sumita and Bhattacharya, Deepta and Blanco-Melo, Daniel and Bloom, David and Boon, Adrianus and Boulant, Steeve and Brandt, Curtis and Broadbent, Andrew and Brooke, Christopher and Cameron, Craig and Campos, Samuel and Caposio, Patrizia and Chan, Gary and Cliffe, Anna and Coffin, John and Collins, Kathleen and Damania, Blossom and Daugherty, Matthew and Debbink, Kari and DeCaprio, James and Dermody, Terence and Dikeakos, Jimmy and DiMaio, Daniel and Dinglasan, Rhoel and Duprex, W. Paul and Dutch, Rebecca and Elde, Nels and Emerman, Michael and Enquist, Lynn and Fane, Bentley and Fernandez-Sesma, Ana and Flenniken, Michelle and Frappier, Lori and Frieman, Matthew and Frueh, Klaus and Gack, Michaela and Gaglia, Marta and Gallagher, Tom and Galloway, Denise and García-Sastre, Adolfo and Geballe, Adam and Glaunsinger, Britt and Goff, Stephen and Greninger, Alexander and Hancock, Meaghan and Harris, Eva and Heaton, Nicholas and Heise, Mark and Heldwein, Ekaterina and Hogue, Brenda and Horner, Stacy and Hutchinson, Edward and Hyser, Joseph and Jackson, William and Kalejta, Robert and Kamil, Jeremy and Karst, Stephanie and Kirchhoff, Frank and Knipe, David and Kowalik, Timothy and Lagunoff, Michael and Laimins, Laimonis and Langlois, Ryan and Lauring, Adam and Lee, Benhur and Leib, David and Liu, Shan-Lu and Longnecker, Richard and Lopez, Carolina and Luftig, Micah and Lund, Jennifer and Manicassamy, Balaji and McFadden, Grant and McIntosh, Michael and Mehle, Andrew and Miller, W. Allen and Mohr, Ian and Moody, Cary and Moorman, Nathaniel and Moscona, Anne and Mounce, Bryan and Munger, Joshua and Münger, Karl and Murphy, Eain and Naghavi, Mojgan and Nelson, Jay and Neufeldt, Christopher and Nikolich, Janko and O'Connor, Christine and Ono, Akira and Orenstein, Walter and Ornelles, David and Ou, Jing-hsiung and Parker, John and Parrish, Colin and Pekosz, Andrew and Pellett, Philip and Pfeiffer, Julie and Plemper, Richard and Polyak, Stephen and Purdy, John and Pyeon, Dohun and Quinones-Mateu, Miguel and Renne, Rolf and Rice, Charles and Schoggins, John and Roller, Richard and Russell, Charles and Sandri-Goldin, Rozanne and Sapp, Martin and Schang, Luis and Schmid, Scott and Schultz-Cherry, Stacey and Semler, Bert and Shenk, Thomas and Silvestri, Guido and Simon, Viviana and Smith, Gregory and Smith, Jason and Spindler, Katherine and Stanifer, Megan and Subbarao, Kanta and Sundquist, Wesley and Suthar, Mehul and Sutton, Troy and Tai, Andrew and Tarakanova, Vera and tenOever, Benjamin and Tibbetts, Scott and Tompkins, Stephen and Toth, Zsolt and van Doorslaer, Koenraad and Vignuzzi, Marco and Wallace, Nicholas and Walsh, Derek and Weekes, Michael and Weinberg, Jason and Weitzman, Matthew and Weller, Sandra and Whelan, Sean and White, Elizabeth and Williams, Bryan and Wobus, Christiane and Wong, Scott and Yurochko, Andrew},
	month = jan,
	year = {2023},
	note = {Publisher: American Society for Microbiology},
	pages = {e00089--23},
}

@article{sandberg_who_2020,
	title = {Who {Should} {We} {Fear} {More}: {Biohackers}, {Disgruntled} {Postdocs}, or {Bad} {Governments}? {A} {Simple} {Risk} {Chain} {Model} of {Biorisk}},
	volume = {18},
	issn = {2326-5108},
	shorttitle = {Who {Should} {We} {Fear} {More}},
	doi = {10.1089/hs.2019.0115},
	language = {eng},
	number = {3},
	journal = {Health Security},
	author = {Sandberg, Anders and Nelson, Cassidy},
	year = {2020},
	pmid = {32522112},
	pmcid = {PMC7310205},
	pages = {155--163},
}

@article{li_machine_2023,
	title = {Machine learning optimization of candidate antibody yields highly diverse sub-nanomolar affinity antibody libraries},
	volume = {14},
	copyright = {2023 The Author(s)},
	issn = {2041-1723},
	url = {https://www.nature.com/articles/s41467-023-39022-2},
	doi = {10.1038/s41467-023-39022-2},
	language = {en},
	number = {1},
	urldate = {2023-08-02},
	journal = {Nature Communications},
	author = {Li, Lin and Gupta, Esther and Spaeth, John and Shing, Leslie and Jaimes, Rafael and Engelhart, Emily and Lopez, Randolph and Caceres, Rajmonda S. and Bepler, Tristan and Walsh, Matthew E.},
	month = jun,
	year = {2023},
	note = {Number: 1
Publisher: Nature Publishing Group},
	pages = {3454},
}

@book{committee_on_strategies_for_identifying_and_addressing_potential_biodefense_vulnerabilities_posed_by_synthetic_biology_biodefense_2018,
	address = {Washington, D.C.},
	title = {Biodefense in the {Age} of {Synthetic} {Biology}},
	isbn = {978-0-309-46518-2},
	url = {https://www.nap.edu/catalog/24890},
	urldate = {2023-01-24},
	publisher = {National Academies Press},
	collaborator = {{Committee on Strategies for Identifying and Addressing Potential Biodefense Vulnerabilities Posed by Synthetic Biology} and {Board on Chemical Sciences and Technology} and {Board on Life Sciences} and {Division on Earth and Life Studies} and {National Academies of Sciences, Engineering, and Medicine}},
	month = dec,
	year = {2018},
	doi = {10.17226/24890},
}

\end{document}